\documentclass{article}

\title{Harmonic potential as an effective limit of a discrete classical interaction}  
\author{Breno R. Segatto\thanks{PIVIC-UFES}, Julio C. S. Azevedo\thanks{PIBIC-CNPq}, 
Manoelito M. de Souza\thanks{e-mail:manoelit@cce.ufes.br }\\
Universidade Federal do Esp\'{\i}rito Santo.
Departamento de F\'{\i}sica \\
29065  Vit\'oria - ES - Brasil}      
\date{January 16, 2003}      

\newcommand{\be}{\begin{equation}}    
\newcommand{\ee}{\end{equation}}

\begin{document}             

\maketitle                   

 
\def \be{\begin{equation}}
\def \ee{\end{equation}}
\def \l{\label}
\medskip
\begin{center}
\end{center}
\par
\bigskip
\begin{abstract}
\noindent Motivated by improving the understanding of the quantum-to-classical transition we use a simple model of classical discrete interactions for studying the discrete-to-continuous transition in the classical harmonic oscillator. A parallel is traced  with gravity  for stressing the relevance of such discrete interaction models.

\end{abstract}
\noindent Fundamental interactions, according to quantum field theory, are realized through the exchange of interaction quanta --- packets of matter-energy with defined quantum numbers, viz. momentum-energy, spin, electric charge, etc. They are discrete interactions, in contradistinction to the classical continuous picture.  The Bohr's correspondence principle, a useful guideline in the early days of quantum mechanics, states that in the limit of very large quantum numbers the classical idea of continuity must result from the quantum discreteness   as an effective description. It would be very interesting to see in a clear way that this discrete-to-continuous transition occurs. This is the objective of the present letter with the use of a simple model of discrete classical interaction for studying this transition in the classical simple harmonic oscillator. We should not forget, however, that the harmonic potential, although being an extremely useful tool in all branches of modern physics, is not itself a fundamental interaction, which, as well known, are just the gravitational, the electromagnetic, the weak and the strong interactions; actually it is just an effective description. This may just valorize the importance of understanding it as an effective limit of a discrete interaction. 
By a classical model of discrete interaction is meant the replacement of the potential, representative of the continuous interaction, by the exchange of, in an evident abuse of language, classical ``\,quanta'', little bits of well defined amounts of energy-momentum. This classical quantum is emitted/absorbed in an instantaneous process caused also by the absorption/emission of a previous quantum. The emitted quantum travels towards its absorber-to-be which is causally reached after a time interval $\Delta t$. Its instantaneous absorption causes an immediate change in the energy-momentum of the absorber and the immediate emission of a new quantum back to the first interaction body. All bodies, as well as their traveling quanta, propagate as free objects in the time interval between consecutive interactions (emission/absorption of quanta). Then, there is no potential energy, only kinetic energy. It is hoped that the discrete-to-continuous transition on such a naive classical model can mimic an enlighten some of the basic aspect of the quantum-to-classical transition on the description of nature.  The search for a better understanding of quantum mechanics started with Einstein, de Broglie, Schrodinger and many others pioneers, including Bohm \cite{Bohm} with his well known proposal, still goes on today with great intensity. Citing 't Hooft who also uses simple classical models with discrete time  \cite{thooft1} in his search for a better understanding of quantum physics ---  ``\,The dividing line between quantum physics and classical physics is more subtle than usually advertised."  \cite{thooft2}. The Khrennikov and Volovich wisdom \cite{Khrennikov} is that the continuous Newton's model is just an approximation of physical reality and that most of the contradiction between the quantum formalism and Newtonian mechanics is, therefore, a consequence of this continuous approximation. This contribution goes along this line of thought.

A non-relativistic regime, which is appropriate for dealing with harmonic oscillator, is assumed. So, we will consider the relative movement of the reduced mass $m$ of a non-relativistic two body-system. At the initial time $t_{0}$ it has the initial position ${\vec{r}}_{0}$ and momentum 
${\vec{p}}_{0}$; it freely propagates on a rectilinear trajectory until its, after then, first interaction  at 
$$t_{1}=t_{0}+\Delta t_{0},$$
$${\vec{r}}_{1}={\vec{r}}_{0}+\Delta{\vec{r}}_{0}={\vec{r}}_{0}+\frac{{\vec{p}}_{0}}{m}\Delta t_{0},$$
that changes its momentum to 
$${\vec{p}}_{1}={\vec{p}}_{0}+\Delta{\vec{p}}_{0}.$$ 
It follows then, again, a rectilinear trajectory until its second interaction at $$t_{2}=t_{1}+\Delta t_{1},$$ which changes its momentum to

$${\vec{p}}_{2}={\vec{p}}_{1}+\Delta{\vec{p}}_{1}.$$
So, at the $n-th$ interaction,
\be
\l{tn}
t_{n}=t_{0}+\sum_{j=0}^{n-1}\Delta t_{j}.
\ee

\be
\l{vpn}
{\vec{p}}_{n}={\vec{p}}_{n-1}+\Delta{\vec{p}}_{n-1}={\vec{p}}_{0}+\sum_{j=0}^{n-1}\Delta{\vec{p}}_{j},
\ee
\be
\l{vrn}
{\vec{r}}_{n}={\vec{r}}_{n-1}+\Delta{\vec{r}}_{n-1}={\vec{r}}_{n-1}+\frac{{\vec{p}}_{n-1}}{m}\Delta t_{n-1}={\vec{r}}_{0}+\frac{1}{m}\sum_{j=0}^{n-1}{\vec{p}}_{j}\Delta t_{j}.
\ee
The position vector, continually changing, describes a polygonal trajectory. There is a sudden change of momentum at each interaction point, a vertex of the polygon. Momentum is clearly a discrete parameter, or at least it changes discretely.
The continuous variables time and position enter in the description of the motion as if they were also discrete parameters only because the interaction events are our reference points for time counting; this is not a lattice calculation. One cannot talk of force or acceleration, just of sudden change of momentum or velocity. They become effective concepts, valid only in appropriate limits, which are our objects of discussion here.

The Eqs. (\ref{vpn}) and (\ref{vrn}) are generic relations valid for any non-relativistic discretely interacting systems.
For the simple harmonic oscillator considered here it is natural to expect that 
\be
\l{hookel}
\Delta{\vec{p}}_{n-1}=-m\omega^2{\vec{r}}_{n-1}\Delta t_{n-1},
\ee
which defines a discrete harmonic oscillator. Among numerously infinite possibilities we will consider the mathematically simplest one
\be
\l{dt}
\Delta t_{n}\equiv\alpha,
\ee
\be
\l{dp}
\Delta{\vec{p}}_{n}\equiv-\alpha m\omega^2{\vec{r}}_{n}.
\ee
$\alpha$ is a positive constant.
The Eqs. (\ref{vpn}) and (\ref{vrn}) become then
\be
\l{vpn1}
{\vec{p}}_{n}={\vec{p}}_{0}-m\alpha\omega^2\sum_{j=0}^{n-1}\;{\vec{r}}_{j},
\ee
\be
\l{vrn1}
{\vec{r}}_{n}=
{\vec{r}}_{0}+\frac{\alpha}{m}\;\sum_{j_{1}=0}^{n-1}{\vec{p}}_{j}.
\ee
The recursive combination of 
Eqs . (\ref{vrn1}) and (\ref{vpn1}) leads  to 
 \be
\l{rns}
{\vec{r}}_{n}=\sum_{s=0}^{[n/2]}(\alpha\;\omega)^{2s}{\vec{r}}_{n}^{\;(s)},
\ee
and
 \be
\l{pns}
{\vec{p}}_{n}=\sum_{s=0}^{[n/2]}(\alpha\;\omega)^{2s}{\vec{p}}_{n}^{\;(s)},
\ee 
where ${\vec{r}}_{n}^{\;(s)}$ and ${\vec{p}}_{n}^{\;(s)}$ are polynomial functions of ${\vec{r}}_{0}$ and ${\vec{p}}_{0},$  $[n/2]$ is the largest integer in $n/2$,
with
\be
\l{rn0}
{\vec{r}}_{n}^{\;(0)}={\vec{r}}_{\;0}+\frac{{\vec{p}}_{0}}{m}\sum_{j_{1}=0}^{n-1}\;\alpha={\vec{r}}_{\;0}+\frac{\alpha\;{\vec{p}}_{0}}{m}n,
\ee
 
\be
{\vec{p}}_{n}^{\;(0)}={\vec{p}}_{0}-m\;\alpha\;\omega^2{\vec{r}}_{0}n,
\ee
\be
\l{drns}
{\vec{r}}_{n}^{\;(s)}=-\sum_{j_{1}=0}^{n-1}\;\sum_{j_{2}=0}^{j_{1}-1}\;{\vec{r}}_{j_{2}}^{\;(s-1)}.
\ee
\be
\l{dpns}
{\vec{p}}_{n}^{\;(s)}=-\sum_{j_{1}=0}^{n-1}\;\sum_{j_{2}=0}^{j_{1}-1}\;{\vec{p}}_{j_{2}}^{\;(s-1)}.
\ee
${\vec{r}}_{n}^{\;(0)}$ would be the position at $t_{n}$ if there were no interaction between $t_{0}$ and $t_{n},$ while ${\vec{p}}_{n}^{\;(0)}$ would be the momentum at $t_{n}$ if all $n$ interactions were equal to the first one;
 the vector ${\vec{r}}_{n}^{\;(1)}$ would be the final position if there were just one interaction in this time interval, and so on. 
Successive re-iterations on Eqs. (\ref{drns}) and (\ref{dpns}) lead to

$$
{\vec{r}}_{n}^{\;(s)}=(-1)^{s}\;\sum_{j_{1}=0}^{n-1}\;\sum_{j_{2}=0}^{j_{1}-1}\dots\sum_{j_{2s}=0}^{j_{2s-1}-1}\;{\vec{r}}_{j_{2s-1}}^{\;(0)}=$$
\be
\l{drns1}
=(-1)^{s}\;\sum_{j_{1}=0}^{n-1}\;\sum_{j_{2}=0}^{j_{1}-1}\dots\sum_{j_{2s}=0}^{j_{2s-1}-1}\;({\vec{r}}_{\;0}+\frac{\alpha{\vec{p}}_{0}}{m}n).
\ee
$$
{\vec{p}}_{n}^{\;(s)}=(-1)^{s}\sum_{j_{1}=0}^{n-1}\sum_{j_{2}=0}^{j_{1}-1}\dots\sum_{j_{2s}=0}^{j_{2s-1}-1}{\vec{p}}_{j_{2s-1}}^{\;(0)}=$$
\be
\l{dpns1}
=(-1)^{s}\;\sum_{j_{1}=0}^{n-1}\;\sum_{j_{2}=0}^{j_{1}-1}\dots\sum_{j_{2s}=0}^{j_{2s-1}-1}\;({\vec{p}}_{0}-m\;\alpha\;\omega^2{\vec{r}}_{0}n).
\ee
Now, we use $n={n\choose 1}$ and the identity
\be
\sum_{j=0}^{n-1}{j\choose k}={n\choose k+1},\qquad{\hbox{for}}\qquad n>k,
\ee
which can easily be proved by induction, to get
\be
\l{rnss}
{\vec{r}}_{n}^{\;(s)}=(-1)^{s}\;({\vec{r}}_{\;0}\;{n\choose 2s
}+\frac{\alpha{\vec{p}}_{0}}{m}{n\choose 2s+1}),
\ee
\be
\l{pnss}
{\vec{p}}_{n}^{\;(s)}=(-1)^{s}\;({\vec{p}}_{\;0}\;{n\choose 2s
}-m\;\alpha\;\omega^2{\vec{r}}_{0}{n\choose 2s+1}),
\ee

which, with Eqs. (\ref{rns}) and (\ref{pns}), leads to
\be
\l{rnf}
{\vec{r}}_{n}=\sum_{s=0}^{[n/2]}(-1)^{s}[(\alpha\omega)^{2s}{\vec{r}}_{\;0}\;{n\choose 2s
}+(\alpha\omega)^{2s+1}\;\frac{{\vec{p}}_{0}}{m\omega}{n\choose 2s+1}],
\ee
\be
\l{vpnf}
{\vec{p}}_{n}=\sum_{s=0}^{[n/2]}(-1)^{s}[(\alpha\omega)^{2s}{\vec{p}}_{\;0}\;{n\choose 2s
}-m\omega(\alpha\omega)^{2s+1}\;{\vec{r}}_{0}{n\choose 2s+1}].
\ee

For $n>>1$, however, the following approximation is valid
\be
\l{nmm1c}
{n\choose k}\approx\frac{n^k}{k!},
\ee
and then 
\be
\l{ccng1}
\sum_{s=0}^{[n/2]}(-1)^{s}(\alpha\omega)^{2s}{n\choose 2s
}\approx\sum_{s=0}^{[n/2]}(-1)^{s}(\alpha\omega)^{2s}\frac{n^{2s}}{(2s)!}\approx\cos(\alpha\omega n),
\ee
\be
\l{scng1}
\sum_{s=0}^{[n/2]}(-1)^{s}(\alpha\omega)^{2s+1}{n\choose 2s+1
}\approx\sum_{s=0}^{[n/2]}(-1)^{s}(\alpha\omega)^{2s+1}\frac{n^{2s+1}}{(2s+1)!}\approx\sin(\alpha\omega n).
\ee
The middle term of Eqs. (\ref{ccng1}) and (\ref{scng1}) are partial sums of the series representing the respective trigonometric functions. Therefore, these trigonometric functions can be seen as the asymptotic limits when $n\rightarrow\infty$ of the respective finite series of combinatorials. Then the Eqs. (\ref{rnf}) and (\ref{vpnf}) become the general solutions of the standard (continuous interaction) harmonic oscillator, valid in this limit as an effective description. For large enough values of the number $n$ of interaction events, the discrete changes of the parameters $t_{n}$, ${\vec{r}}_{n}$ and ${\vec{p}}_{n}$ are so small that they can effectively be described as if they were continuously changing under the action of a continuous interaction field, the harmonic potential. In this limit
\be
\l{Hn}
H_{n}\equiv\frac{1}{2m}{\vec{p}}_{n}^{\;2}+\frac{\alpha\;\omega^2}{2}{\vec{r}}_{n}^{\;2}=H_{0}.
\ee
Similar or related behavior have been proved for classical electrodynamics \cite{1}, Newtonian gravitation \cite{2}, general relativity \cite{3} and, generically, for field theory \cite{4}.

The harmonic potential appears as an effective continuous interaction with two asymptotic (in the limit of very large $n$) conditions:
\be
\l{nmm1}
n>>1
\ee
and
\be
\l{awnmm1}
\alpha\;\omega n=\omega (t_{n}-t_{0})<<1,
\ee
for the trigonometric function retrieval. We made use of $t_{n}=t_{0}+\alpha n$. In such a limit we can safely treat changes due to $\Delta n=1<<n$ as time derivatives
\be
\frac{d}{dt_{n}}=\frac{dn}{dt_{n}}\frac{d}{dn}=\frac{1}{\alpha}\frac{d}{dn}.
\ee

For keeping track of the order of what is being neglected with the approximation (\ref{nmm1c}) that leads to the continuous description we use $\delta ^{(k)}$ for indicating the $k$-th largest contribution in a given neglected term.
\be
\delta^{(1)} {n\choose k}=-\frac{n^{k-1}}{k!}{k\choose 2}=-\frac{1}{2}\frac{n^{k-1}}{(k-2)!}.
\ee
Then, from  Eqs. (\ref{rnf}) and (\ref{vpnf})
\be
\l{rnf1}
\delta^{(1)}{\vec{r}}_{n}=\sum_{s=0}^{[n/2]}(-1)^{s}[(\alpha\;\omega)^{2s}{\vec{r}}_{\;0}\;\delta^{(1)}{n\choose 2s
}+(\alpha\omega)^{2s+1}\;\frac{{\vec{p}}_{0}}{m\;\omega}\delta^{(1)}{n\choose 2s+1}],
\ee
\be
\l{vpnf1}
\delta^{(1)}{\vec{p}}_{n}=\sum_{s=0}^{[n/2]}(-1)^{s}[(\alpha\omega)^{2s}{\vec{p}}_{\;0}\;\delta^{(1)}{n\choose 2s
}-m\;\omega(\alpha\omega)^{2s+1}\;{\vec{r}}_{0}\delta^{(1)}{n\choose 2s+1}],
\ee
from which we get

\be
\delta^{(1)}{\vec{r}}_{n}=-(\alpha\omega)^2\frac{n}{2}\;{\vec{r}}_{n},\qquad \delta^{(1)}{\vec{p}}_{n}=-(\alpha\omega)^2\frac{n}{2}\;{\vec{p}}_{n},
\ee
so that
\be
\delta^{(1)}H_{n}\approx H_{n}(\frac{{\vec{p}}_{n}\delta^{(1)}{\vec{p}}_{n}}{m}+m\omega^{2}{\vec{r}}_{n}\delta^{(1)}{\vec{r}}_{n})=-(\alpha\omega)^2 n\;H_{n}.
\ee
The fractional neglected terms, according to Eq. (\ref{awnmm1}), are of the order of $\frac{1}{n}<<1$. Whereas the effective continuous limit is not affected by the choice made in Eq. (\ref{dt}), the neglected contributions $\delta^{(1)}$ are strongly affected. It is a characteristic of discrete interaction models that first order corrections ($\delta^{(1)}$) to the effective continuous fields are proportional to $n$ (or $t_{n}$). This may be not very relevant for the harmonic oscillator or any system with periodic motion, in contradistinction to, for example, a nearly radial escaping motion\footnote{The anomalous acceleration observed in the Pioneer effect \cite{Pioneer}, not detected in planetary motion, can be an example.} in a gravitational field\footnote{Shield effects prevent  this happening with electromagnetic interactions, the other long-range fundamental interaction.}. In this case, a large/small $n$ (or $t_{n}$) implies a large/small $r_{n}$. There are then three distinct scales to be considered:
\begin{enumerate}
\item$n$ not large enough for the validity of the continuous interaction approximation. Necessarily, a discrete interaction description must be used. This corresponds to microscopic distances, a domain still lacking experimental data on gravity \cite{5}. 
\item Very large $n$ that validates the effective continuous field description, but not large enough so that the $\delta^{(1)}$-corrections may be neglected. This corresponds to the solar system size scale where the continuous (Newton's and Einstein's) descriptions have solid experimental confirmation \cite{6}.
\item Values of $n$ so large that the $\delta^{(1)}$-corrections cannot be neglected. It is well known that for galactic and cosmological distance scales the standard continuous descriptions of gravity require the ad hoc assumption of, up-to-now, unseen new forms of (dark) matter and energy \cite{7}. These new actors are called for compensating the inability of the standard field to produce a repulsive interaction component (this {\it{wrong}}-sign interaction is necessary for fitting the data) from their known sources\cite{Manheim}. The $\delta^{(1)}$-corrections terms can do that \cite{2} and provide a natural interpretation for this interaction {\it{wrong}}-sign. This, however, is in direct contradiction to the predictions of quantum gravity for which effects of discrete gravity are to be revealed only at ultra high energies or ultra short distances.
\end{enumerate}

\bigskip

\begin{center}{Acknowledgements}
\end{center}
Azevedo, J.C.S. and Segatto, B.R. acknowledge their CNPq grants for undergraduate studies.

 \end{document}